\documentclass{nature}

\usepackage{amsmath}
\usepackage{amssymb}
\usepackage{mathrsfs}
\usepackage{paralist}
\usepackage{hyperref}
\usepackage{booktabs}
\usepackage{graphicx}
\usepackage{float}
\usepackage{morefloats}
\usepackage{multirow}
\usepackage{color}
\usepackage{paralist}
\usepackage{todonotes}
\usepackage{wrapfig}
\usepackage{threeparttable}
\usepackage{colortbl}
\usepackage{bbding}
\definecolor{bblue}{rgb}{0,150,230}
\definecolor{mygray}{gray}{.9}
\definecolor{myy}{RGB}{126,95,0}

\usepackage{bm}

\newcommand{\eg}[1]{\textit{e.g.,}}
\newcommand{\ie}[1]{\textit{i.e.,}}

\newcommand{\figref}[1]{Fig.\!~\ref{#1}}

\usepackage{cleveref}
\crefname{section}{}{§§}
\Crefname{section}{}{§§}

\usepackage{caption}
\newcommand\blfootnote[1] 
{
  \begingroup
  
  \footnotetext{#1}
  \addtocounter{footnote}{-1}
  \endgroup
}

\title{Deep multi-modal aggregation network for MR image reconstruction with auxiliary modality}

\author{Chun-Mei Feng\textsuperscript{1}, Yong Xu\textsuperscript{1,\Envelope}, 
Ling Shao\textsuperscript{2}, 
David Zhang\textsuperscript{3,4,5}, 
and Huazhu Fu\textsuperscript{6, \Envelope}}
\begin{document}

\maketitle

\begin{affiliations}
 \item Shenzhen Key Laboratory of Visual Object Detection and Recognition, Harbin Institute of Technology (Shenzhen), 518055, China.
 \item National Center for Artificial Intelligence (NCAI), SDAIA, KSA.
 \item School of Science and Engineering, The Chinese University of Hong Kong (Shenzhen), Shenzhen 518172, China. 
 \item Shenzhen Research Institute of Big Data, Shenzhen 518172, China. \item Shenzhen Institute of Artificial Intelligence and Robotics for Society, Shenzhen 518172, China.
 \item Institute of High Performance Computing, Agency for Science, Technology and Research, Singapore 138632.
\end{affiliations}

\begin{abstract}
Magnetic resonance (MR) imaging produces detailed images of organs and tissues with better contrast, but it suffers from a long acquisition time, which makes the image quality vulnerable to say motion artifacts. Recently, many approaches have been developed to reconstruct full-sampled images from partially observed measurements to accelerate MR imaging. However, most approaches focused on reconstruction over a single modality, neglecting the discovery of correlation knowledge between the different modalities. Here we propose a Multi-modal Aggregation network for mR Image recOnstruction with auxiliary modality (MARIO), which is capable of discovering complementary representations from a fully sampled auxiliary modality, with which to hierarchically guide the reconstruction of a given target modality. This implies that our method can selectively aggregate multi-modal representations for better reconstruction, yielding comprehensive, multi-scale, multi-modal feature fusion. Extensive experiments on IXI and fastMRI datasets demonstrate the superiority of the proposed approach over state-of-the-art MR image reconstruction methods in removing artifacts.
\end{abstract}

\blfootnote{\noindent \Envelope~Corresponding author: Yong Xu and Huazhu Fu (\texttt{yongxu@ymail.com, hzfu@ieee.org})}

\newpage

\section*{Introduction}
\label{sec1}
Magnetic resonance (MR) imaging is a widespread scanning technique for musculoskeletal, neurological, and oncological diseases\cite{kaittanis2014environment,evans2020non}. The raw data obtained by the scanner is complex-valued data, collected in $k$-space and transferred to the image domain by inverse 2D Fourier transform\cite{moratal2008k}. However, the physical nature of the MR data acquisition process makes the scanning time up to tens of minutes long, causing uncomfortable examination experiences and high health care costs\cite{Caliva2022}. Therefore, accelerated MR imaging is a major ongoing research goal towards improving the patient experience. The acceleration of MR imaging can be addressed by prospective or retrospective techniques\cite{liu2021learning,heijman2007comparison,kober2012prospective}. The former uses optical tracking of a target marker placed on the head or images continuously reacquired from a dedicated navigator scan for real-time motion prediction\cite{zaitsev2006magnetic,maclaren2013prospective,qin2009prospective}. This technology requires additional hardware material support, resulting in expensive economic consumption. The latter can be either acquisition-based or software-based. For example, the classic acquisition-based method is PROPELLER\cite{dogdas2010motion}, which uses self-navigation technology to repeatedly cover the $k$-space with concentric blades rotating at different angles. However, because of the specific navigator sequence, this strategy can increase acquisition time and limit imaging parameters, \ie, repetition/echo/inversion time\cite{liu2021learning}.
The software-based method reconstructs images from under-sampled $k$-space measurements to achieve fast MR imaging without modifying sequence installation markers and constrained acquisition parameters. This is an economical, quick, and easy-to-implement post-processing approach. However, obtaining a basic reconstruction of zero-filled $k$-space data often exhibit aliasing artifacts, making them inappropriate for clinical diagnosis\cite{moratal2008k,hu2020brain}. Therefore, how to reduce these artifacts and recover high-fidelity images from the insufficient $k$-space measurements becomes the goal of MR image reconstruction systems. Relying on the inherently redundant in $k$-space\cite{xiang2017deep}, compressed sensing (CS) and parallel imaging have made significant progress in MR image reconstruction\cite{lustig2007sparse,pruessmann1999sense,griswold2002generalized}. 
However, CS-based methods for MR image reconstruction require significant time for iterative minimization, resulting in difficulty in near-real-time MR image scenarios, \eg, cardio-MR and functional-MR imaging\cite{yu2017deep}. Parallel imaging reconstructs the images from multi-channel aliased images using the coil sensitivity profiles through multiple-channels, but it will cause artifacts if the in-plane acceleration rate is higher than 2\cite{hutchinson1988fast}.

Recently, deep learning techniques have been widely used for MR image reconstruction\cite{wang_first,yang2017dagan,quan2018compressed,sriram2020grappanet,Wang2020nmi,Caliva2022}. However, most of these methods focus on using undersampled data of a mono-modal acquisition to reconstruct high-quality MR images. In clinical practice, collecting multiply modalities at the same time is often required to better observe the lesions. Therefore, using MR sequences with a shorter acquisition time to assist the collection of other sequences closely related to structural information will be an efficient solution to shorten the overall imaging time. For example, T1 and T2 weighted images are two closely related MR sequences, but T1 weighted imaging is slower than T2 weighted due to its relatively longer TRs and TEs\cite{sui2018multimodal}. Therefore, we can use the T1 weight as an auxiliary modality to guide T2 weighted image reconstruction. 
Currently, only a few studies have attempted to use deep learning techniques for multi-modal fusion between different MR sequences to accelerate image reconstruction\cite{xiang2018deep}, \eg, reconstruct the high-quality T2 weighted image according to the high coupling relationship between T1 and T2 images\cite{xiang2018deep, zhou2020dudornet}. However, these methods stacking T1 image and T2 image directly into the deep network do not explore the feature level information, which is especially critical for T2 weighted image reconstruction.

%

In this study, we develop a Multi-modal Aggregation network for mR Image recOnstruction with auxiliary modality (MARIO), to explore the hierarchical correlation between different modalities.
Unlike previous methods that only take T1 weights as priors, our MARIO enhances the learning of the T1 stream at multiple scales with T1-guided attention modules. Specifically, we design a two-stream network to learn a specific representation for each modality. The attention module fuses the T1 weighted features in each convolutional stage to learn effective feature representations. The T1 features are then enhanced to aid in the learning of T2 features. Finally, benefiting from these modality-specific networks and fusion module, our MARIO can learn a latent representation in a complementary way to generate the target images.

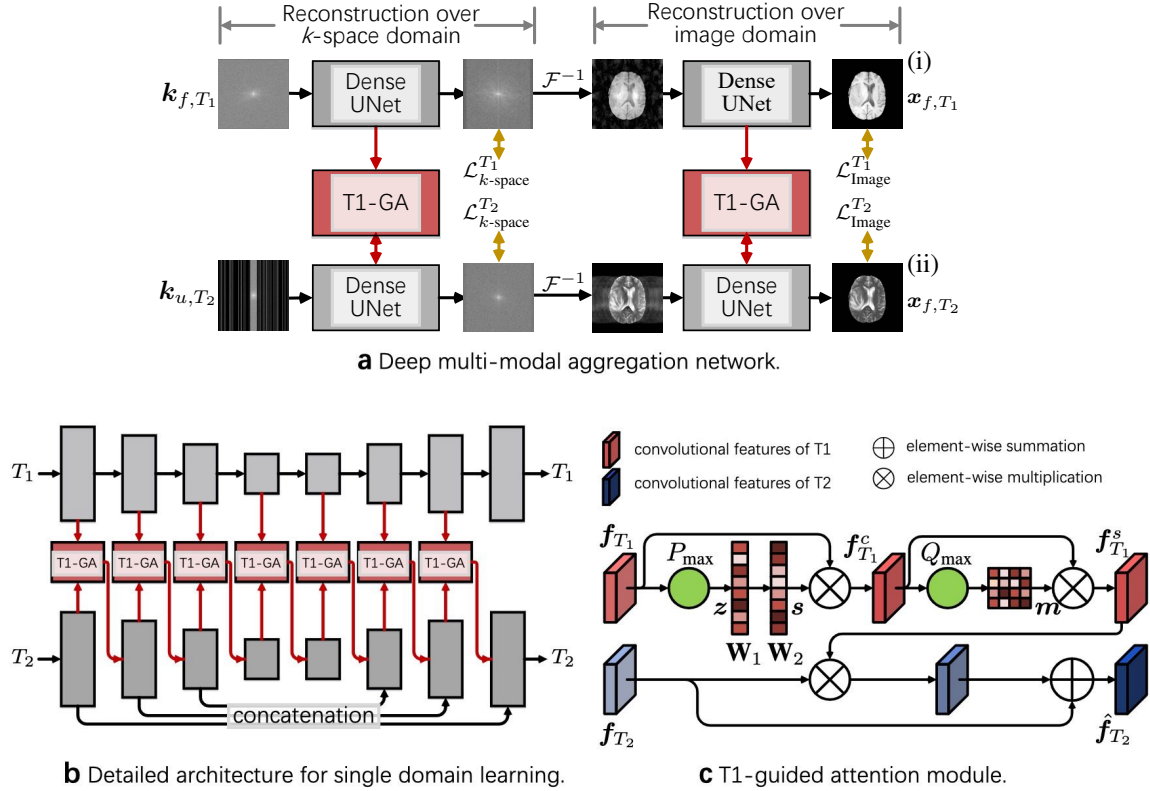
\begin{figure*}[!t]
\centering
  \includegraphics[width=0.92\textwidth]{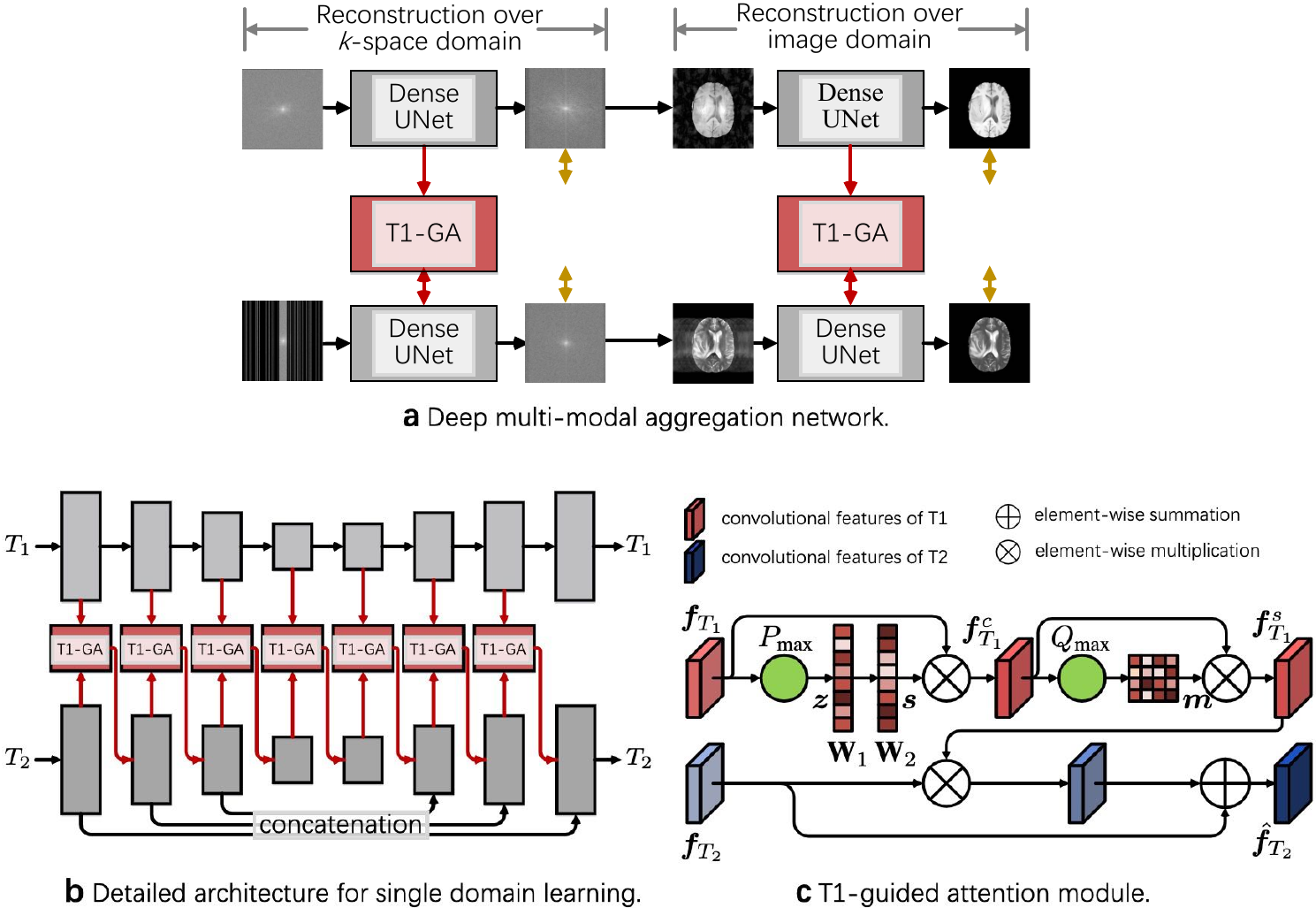}
  \put(-92,272){\footnotesize (i)}
  \put(-92,195){\footnotesize (ii)}
  \put(-374,260){\footnotesize $\bm{k}_{f,T_1}$}
  \put(-375,185){\footnotesize $\bm{k}_{u,T_2}$}
  \put(-260,231){\scriptsize $\mathcal{L}_{k\text{-space}}^{T_1}$}
  \put(-260,215){\scriptsize $\mathcal{L}_{k\text{-space}}^{T_2}$}
  \put(-119,231){\scriptsize $\mathcal{L}_{\text{Image}}^{T_1}$}
  \put(-119,215){\scriptsize $\mathcal{L}_{\text{Image}}^{T_2}$}
  \put(-230,265){\scriptsize $\mathcal{F}^{-1\!}$}
  \put(-230,188){\scriptsize $\mathcal{F}^{-1\!}$}
  \put(-92,259){\scriptsize $\bm{x}_{f, T_1}$}
  \put(-92,182){\scriptsize $\bm{x}_{f, T_2}$}
  \caption{\textbf{Overview of MARIO.} \textbf{a}) Illustration of the proposed MARIO, including two inputs, \ie, fully sampled T1 weighted $k$-space of the frequency domain and undersampled T2 weighted $k$-space of the frequency domain, and four losses, \ie, $\mathcal{L}_{k\text{-space}}^{T_1}$ and $\mathcal{L}_{k\text{-space}}^{T_2}$ over the $k$-space domain reconstruction, and $\mathcal{L}_{\text{Image}}^{T_1}$, $\mathcal{L}_{\text{Image}}^{T_2}$ over the image domain reconstruction. \textbf{b}) The detailed network architecture for learning in an individual domain, where the T1 branch are used to assist the the T2 branch by the T1-GA module in each layer. \textbf{c}) Architecture of the T1-guided attention module (T1-GA) for multi-modal feature fusion. $\bm{f}_{T_1}$ and $\bm{f}_{T_2}$ are the input features of the two modalities, $\bm{f}_{T_1}^s$ and $\hat{\bm{f}}_{T_2}$ are the output features which contains rich complementary information.}
  \label{T1T2}
\end{figure*}

\section*{Results}
\subsection{MARIO: Deep multi-modal aggregation network for exploring multi-scale hierarchical features.} Our MARIO is a multi-modal aggregation process that excavates the multi-scale properties of an auxiliary modality to reconstruct the target modality. It considers the complementary and diversity of different MR modalities.
An overview of MARIO is shown in~\figref{T1T2}, whose goal is to fully explore the relationships between multi-modal MR images and thereby accelerate MR imaging with complementary of the auxiliary modality. MARIO adopts Dense-UNets structure in a hybrid domain, firstly restores the $k$-space signal in frequency domain, and then restores the textures in the image domain.
Further, a novel T1-guided attention module is proposed to enhance the features from the T1 weight, which provides a robust fusion method to identify the latent features specific to the target modality and effectively boosts the fusion performance. It is worth noting that MARIO can deeply mine auxiliary modality layer by layer at different scales as supplementary information for the target modality reconstruction.

\subsection{T2 weighted image reconstruction with the guidance of T1 weighted feature.}
\begin{figure*}[!t]
\centering
  \includegraphics[width=1\textwidth]{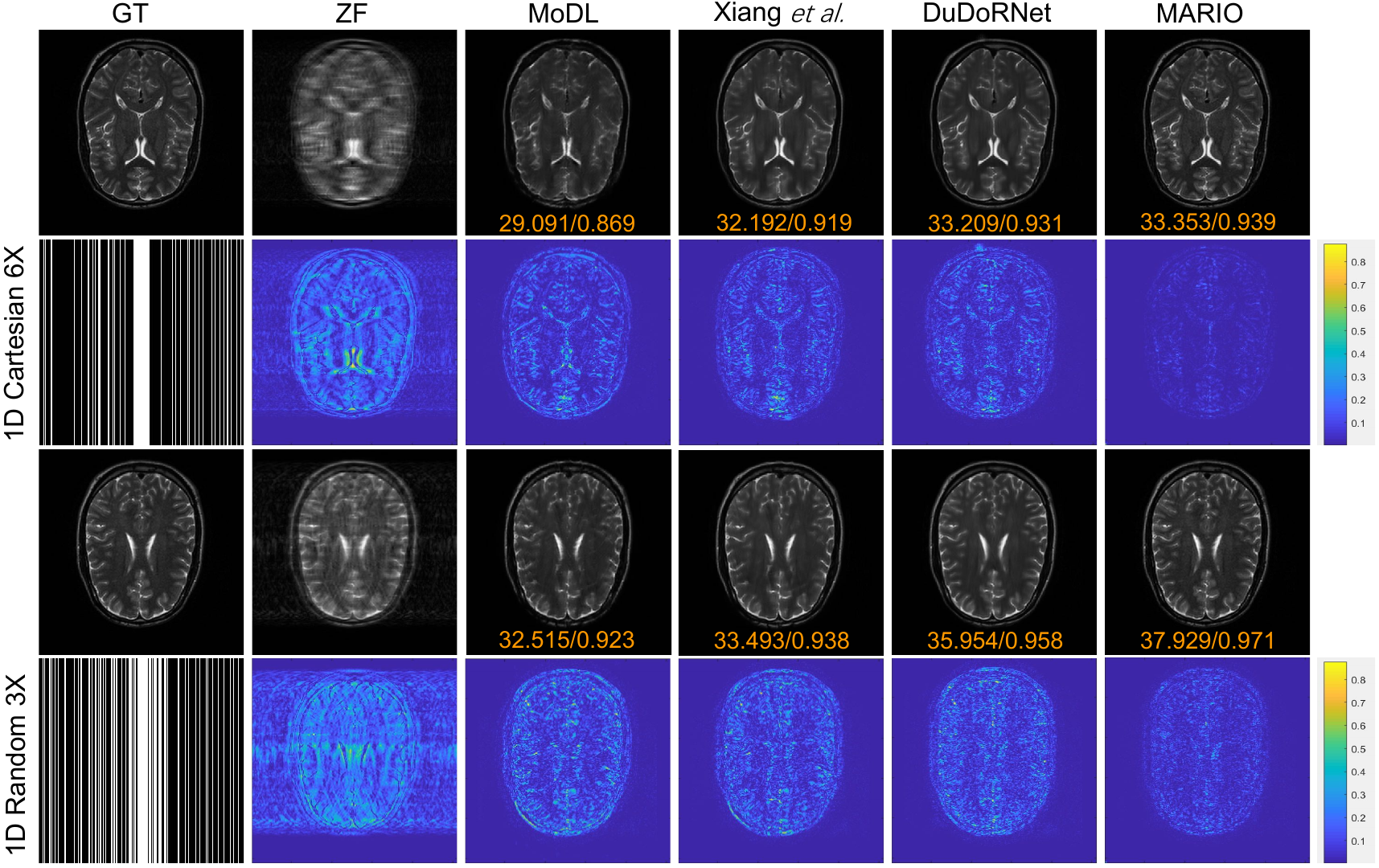}
  \caption{\textbf{Visual comparison of different methods under 1D undersampling patterns with \textbf{3$\times$} and \textbf{6$\times$} acceleration rate.} Reconstruction results and error maps are presented with corresponding quantitative measurements in PSNR/SSIM. The less texture in the error map, the better the image quality.}
  \label{1D}
\end{figure*}

\begin{figure*}[!t]
\centering
  \includegraphics[width=1\textwidth]{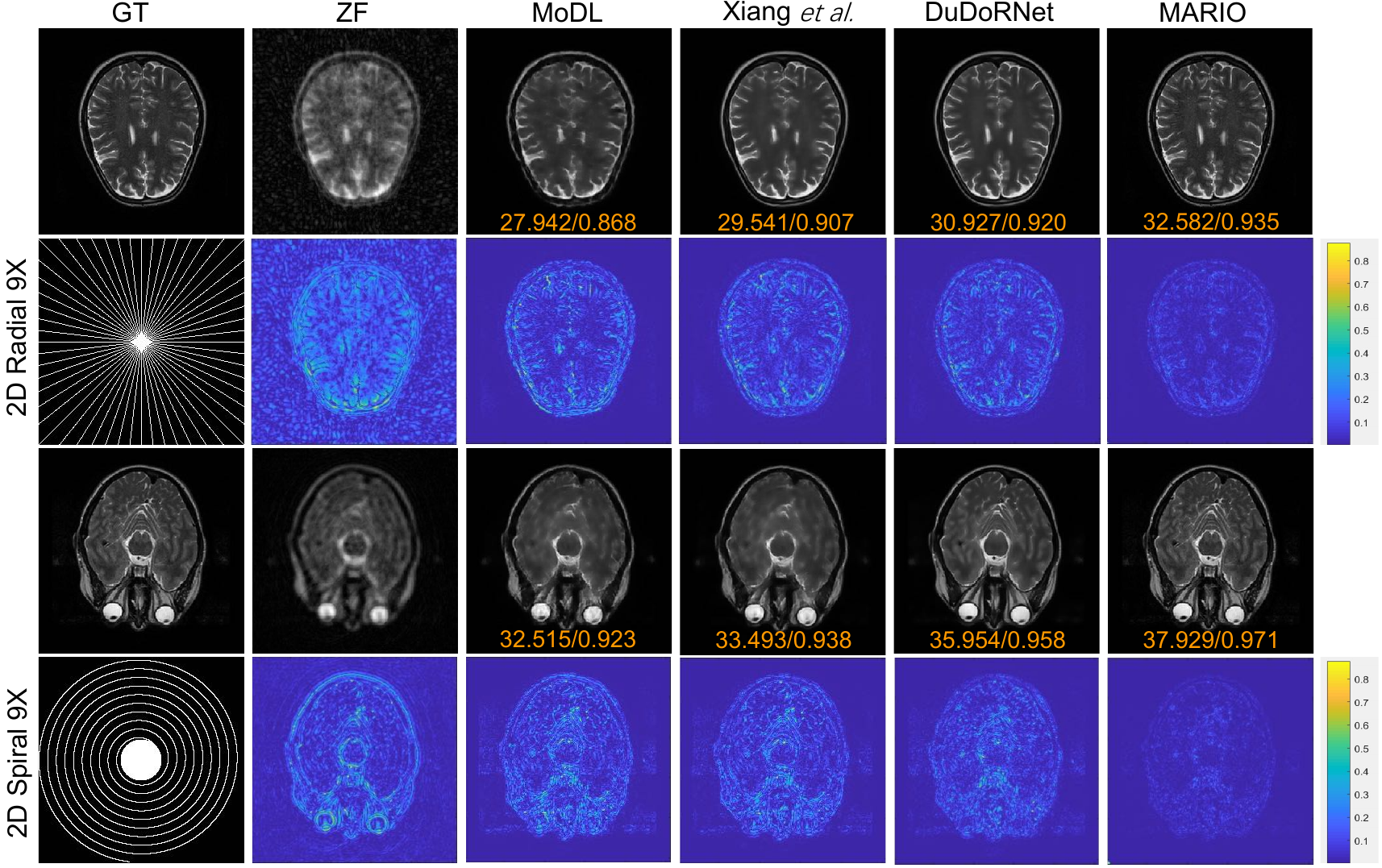}
  \caption{\textbf{Visual comparison of different methods under \textbf{2D undersampling patterns} with \textbf{9$\times$} acceleration rate.} Reconstruction results and error maps are presented with corresponding quantitative measurements in PSNR/SSIM. The less texture in the error map, the better the image quality.}
  \label{2D}
\end{figure*}

We first evaluate the reconstruction of T2 weighted images under the guidance of T1 weighted feature. The IXI dataset\cite{IXIDataset} is used, which consists of 577 subjects with different MR sequences from three different hospitals in London (\ie, Hammersmith Hospital using a Philips 3T system, Guy’s Hospital using a Philips 1.5T system, and Institute of Psychiatry using a GE 1.5T system). In this study, we use the T1 weight images in this dataset to guide T2 image reconstruction. We randomly split the data into 369 subjects for training, 93 subjects for validation, and 115 subjects for testing. Each subject with paired T1 and T2 weighted images and the volume is cropped to a size of 256$\times$256$\times$100. For the IXI dataset, the model is trained on the single-coil magnitude images. 
We compare our MARIO with several state-of-the-art methods that are closely related to our task, \ie, MoDL\cite{aggarwal2018modl}, the classical model-based deep learning, on single-modal MR image data, a multi-modal fusion method proposed by Xiang \textit{et al.}\cite{xiang2018deep}, and DuDoRNet\cite{zhou2020dudornet}. All these methods are trained on the single-coil magnitude images with the same setting. For fair comparison, Xiang \textit{et al.}\cite{xiang2018deep} and DuDoRNet\cite{zhou2020dudornet} are retrained using their default parameter settings, where the number of feature maps in Xiang \textit{et al.}\cite{xiang2018deep} is 64. The superiority of the proposed method is validated by various experiment settings. For IXI dataset, we use four different undersampling masks with different acceleration rates to obtain the various undersampled measurements, including 1D Uniform, 1D Random, 2D Radial, and 2D Spiral patterns with 3$\times$, 6$\times$, and 9$\times$. Rician noise is also randomly generated with a power of 10$\%$ of mean voxel intensities and added to the $k$-space data before applying the undersampling. Examples of the undersampling patterns are illustrated in the left columns in~\figref{1D} and~\figref{2D}.

The experiment results are shown in the first four sub-tables in Table~\ref{table:1}. Each reports the average SSIM and PSNR scores with respect to different undersampling patterns and acceleration factors. The first two sub-tables summarize the image quality under 1D Uniform and 1D Random sampling patterns with different acceleration rates. The second two sub-tables summarize the image quality under 2D Radial and 2D Spiral sampling patterns with different acceleration rates. We find that our MARIO with these undersampling patterns yields the best results among all acceleration rate and reconstruction methods. This suggests that our model can effectively fuse the T1 weight features from the complementary modality and that this is beneficial to the reconstruction of the target modality. Specifically, without the guidance of T1 weight data, the restoration of the T2 weight image is far less effective than when using other multi-modal fusion models. 
Although Xiang \textit{et al.}\cite{xiang2018deep} also add the T1 weight information into the model, it directly concatenates the T1 and T2 weights together as the input of the network, rather than explicitly learning the characteristics of T1 weight to help the T2 reconstruction. 
Therefore, its performance is less effective in comparison with other fusion models. In particular, under the 1D Random sampling mask with 3$\times$ acceleration, we improve the PSNR from 33.929 dB to 36.927 dB, and SSIM from 0.945 to 0.969, as compared to the current best approach, \ie, DuDoRNet. Since DuDoRNet and Xiang \textit{et al.} only use the auxiliary modality as prior information, without any changes in the network, they have fewer parameters than our method. However, their reconstruction results are inferior to those produced by our model. Additionally, it is obvious that 1D undersampling patterns make the image more difficult to restore than when using 2D patterns, and the reconstruction becomes more difficult when the acceleration rate increases. Under 2D Radial undersampling patterns, our MARIO achieves scores of 34.317 and 32.872 in terms of PSNR at 6$\times$ and 9$\times$ acceleration rates, respectively. Moreover, our model also achieves compelling performance under the 2D Spiral patterns.

\begin{figure*}[!t]

	\makeatletter\def\@captype{table}\makeatother\caption{\textbf{ Average (with standard deviation) results under different undersampling patterns, in terms of SSIM, and PSNR.} The best and second-best results are marked in {\color{red}red} and {\color{blue}blue}, respectively. Compared with the other methods, MARIO yields the highest PSNR and SSIM results.} 
	\begin{minipage}[t]{0.5\textwidth}
		\centering
        \label{table:1}
		\begin{threeparttable}
			\resizebox{1\textwidth}{!}{
				\setlength\tabcolsep{4pt}
				\renewcommand\arraystretch{1.1}
				\begin{tabular}{r||cc|cc}
			        \hline\hline
			        \rowcolor{mygray}
			        &  \multicolumn{2}{c|}{1D-Uniform-6$\times$} &  \multicolumn{2}{c}{1D-Uniform-9$\times$} \\
			        \rowcolor{mygray}

			        \multirow{-2}{*}{IXI} &~~SSIM$\uparrow$~~ & PSNR$\uparrow$  &~~SSIM$\uparrow$~~ & PSNR$\uparrow$ \\ \hline\hline
                    ZF  &0.756$\pm$0.06 &25.9$\pm$1.93 &0.703$\pm$0.08 &25.9$\pm$2.11 \\ 
                    MoDL &0.897$\pm$0.03 &29.5$\pm$1.01 &0.880$\pm$0.05 &28.3$\pm$1.01\\ 
                    Xiang \textit{et al.} &0.935$\pm$0.04 &31.5$\pm$1.52 &0.955$\pm$0.08 &31.0$\pm$1.47 \\ 
                    DuDoRNet &{\color{blue}0.937$\pm$0.04} &{\color{blue}32.6$\pm$1.72} &{\color{blue}0.930$\pm$0.08} &{\color{blue}31.8$\pm$1.38} \\ \hline
                    \textbf{MARIO}  &{\color{red}0.944$\pm$0.02} &{\color{red}32.9$\pm$1.02} &{\color{red}0.947$\pm$0.04} &{\color{red}32.4$\pm$1.16} \\ \hline
	        	\end{tabular}
	        	}

		\end{threeparttable}
       \vspace*{2pt}
	\end{minipage}
	\begin{minipage}[t]{0.5\textwidth}
		\centering
		\begin{threeparttable}
			\resizebox{1\textwidth}{!}{
				\setlength\tabcolsep{4pt}
				\renewcommand\arraystretch{1.1}
				\begin{tabular}{r||cc|cc}
			        \hline\hline
			        \rowcolor{mygray}
			        &  \multicolumn{2}{c|}{1D-Random-3$\times$} &  \multicolumn{2}{c}{1D-Random-6$\times$} \\
			        \rowcolor{mygray}

			        \multirow{-2}{*}{IXI} &~~SSIM$\uparrow$~~ & PSNR$\uparrow$ &~~SSIM$\uparrow$~~ & PSNR$\uparrow$ \\ \hline\hline
                    ZF &0.747$\pm$0.10 &27.8$\pm$1.73 &0.694$\pm$0.12 &25.9$\pm$2.28 \\ 
                    MoDL &0.926$\pm$0.14 &32.6$\pm$1.02 &0.882$\pm$0.07 &29.3$\pm$1.28\\ 
                    Xiang \textit{et al.} &0.932$\pm$0.08 &32.5$\pm$1.21 &0.920$\pm$0.04 &31.6$\pm$1.09 \\ 
                    DuDoRNet &{\color{blue}0.945$\pm$0.07} &{\color{blue}33.9$\pm$1.09} &{\color{blue}0.926$\pm$0.07} &{\color{blue}31.9$\pm$1.29} \\ \hline
                    \textbf{MARIO}  &{\color{red}0.969$\pm$0.06} &{\color{red}36.9$\pm$1.18} &{\color{red}0.949$\pm$0.04} &{\color{red}33.1$\pm$1.02} \\ \hline
	        	\end{tabular}
			}
		\end{threeparttable}
       \vspace*{2pt}

	\end{minipage}

		\begin{minipage}[t]{0.5\textwidth}
		\centering
		\begin{threeparttable}
			\resizebox{1\textwidth}{!}{
				\setlength\tabcolsep{4pt}
				\renewcommand\arraystretch{1.1}
				\begin{tabular}{r||cc|cc}
			        \hline\hline
			        \rowcolor{mygray}
			        &  \multicolumn{2}{c|}{2D-Radial-6$\times$} &  \multicolumn{2}{c}{2D-Radial-9$\times$} \\
			        \rowcolor{mygray}

			        \multirow{-2}{*}{IXI} &~~SSIM$\uparrow$~~ & PSNR$\uparrow$ &~~SSIM$\uparrow$~~ & PSNR$\uparrow$ \\ \hline\hline
                    ZF &0.487$\pm$0.08 &27.1$\pm$1.86 &0.414$\pm$0.07 &25.2$\pm$1.28 \\ 
                    MoDL &0.894$\pm$0.06 &31.3$\pm$1.09 &0.853$\pm$0.06 &28.8$\pm$1.17\\ 
                    Xiang \textit{et al.} &0.905$\pm$0.08 &32.3$\pm$1.68 &0.890$\pm$0.08 &30.5$\pm$1.72 \\ 
                    DuDoRNet &{\color{blue}0.931$\pm$0.06} &{\color{blue}33.3$\pm$1.02} &{\color{blue}0.912$\pm$0.06} &{\color{blue}31.5$\pm$1.27} \\ \hline
                    \textbf{MARIO}  &{\color{red}0.947$\pm$0.07} &{\color{red}34.3$\pm$1.31} &{\color{red}0.927$\pm$0.05} &{\color{red}32.9$\pm$1.11} \\ \hline
	        	\end{tabular}
			}
		\end{threeparttable}
       \vspace*{2pt}
	\end{minipage}
	\begin{minipage}[t]{0.5\textwidth}
		\centering
		\begin{threeparttable}
			\resizebox{1\textwidth}{!}{
				\setlength\tabcolsep{4pt}
				\renewcommand\arraystretch{1.1}
				\begin{tabular}{r||cc|cc}
			        \hline\hline
			        \rowcolor{mygray}
			        &  \multicolumn{2}{c|}{2D-Spiral-6$\times$} &  \multicolumn{2}{c}{2D-Spiral-9$\times$} \\
			        \rowcolor{mygray}

			        \multirow{-2}{*}{IXI} &~~SSIM$\uparrow$~~ & PSNR$\uparrow$ &~~SSIM$\uparrow$~~ & PSNR$\uparrow$ \\ \hline\hline
                    ZF &0.617$\pm$0.06 &28.1$\pm$1.28 &0.638$\pm$0.07 &27.6$\pm$2.12 \\ 
                    MoDL &0.894$\pm$0.06 &31.2$\pm$1.42 &0.883$\pm$0.04 &30.4$\pm$1.08\\ 
                    Xiang \textit{et al.} &0.905$\pm$0.05 &31.8$\pm$1.07 &0.890$\pm$0.03 &30.6$\pm$1.21 \\     
                    DuDoRNet &{\color{blue}0.942$\pm$0.04} &{\color{blue}34.1$\pm$1.27} &{\color{blue}0.939$\pm$0.05} &{\color{red}33.1$\pm$1.65} \\ \hline
                    \textbf{MARIO}  &{\color{red}0.951$\pm$0.04} &{\color{red}34.5$\pm$1.17} &{\color{red}0.940$\pm$0.03} &{\color{blue}33.0$\pm$1.20} \\ \hline
	        	\end{tabular}
			}
		\end{threeparttable}
       \vspace*{2pt}
	\end{minipage}

	\begin{minipage}[t]{0.5\textwidth}
		\centering
		\vspace{2mm}
		\begin{threeparttable}
			\resizebox{1\textwidth}{!}{
				\setlength\tabcolsep{4pt}
				\renewcommand\arraystretch{1.1}
				\begin{tabular}{r||cc|cc}
			        \hline\hline
			        \rowcolor{mygray}
			        &  \multicolumn{2}{c|}{Random-4$\times$} &  \multicolumn{2}{c}{Random-8$\times$} \\
			        \rowcolor{mygray}

			        \multirow{-2}{*}{fastMRI} &~~SSIM$\uparrow$~~ & PSNR$\uparrow$  &~~SSIM$\uparrow$~~ & PSNR$\uparrow$ \\ \hline\hline
                    MoDL-PDFS
                    &0.630$\pm$0.04 &29.2$\pm$1.55
                    &0.550$\pm$0.02 &28.1$\pm$1.07 \\ 
                    Xiang \textit{et al.} &0.600$\pm$0.05 &29.3$\pm$1.54 &0.562$\pm$0.06 &28.4$\pm$1.25 \\ 
                    DuDoRNet &{\color{blue}0.640$\pm$0.02} &{\color{blue}29.5$\pm$1.33} &{\color{blue}0.571$\pm$0.05} &{\color{blue}28.7$\pm$1.38} \\ \hline
                    \textbf{MARIO}  &{\color{red}0.652$\pm$0.02} &{\color{red}30.3$\pm$1.12} &{\color{red}0.604$\pm$0.05} &{\color{red}29.4$\pm$1.09} \\ \hline
	        	\end{tabular}
			}

		\end{threeparttable}
		\vspace*{2pt}
	\end{minipage}
	\begin{minipage}[t]{0.5\textwidth}
		\centering
		\vspace{2mm}

		\begin{threeparttable}
			\resizebox{1\textwidth}{!}{
				\setlength\tabcolsep{4pt}
				\renewcommand\arraystretch{1.1}
				\begin{tabular}{r||cc|cc}
			        \hline\hline
			        \rowcolor{mygray}
			        &  \multicolumn{2}{c|}{Equispaced-4$\times$} &  \multicolumn{2}{c}{Equispaced-8$\times$} \\
			        \rowcolor{mygray}

			        \multirow{-2}{*}{fastMRI} &~~SSIM$\uparrow$~~ & PSNR$\uparrow$ &~~SSIM$\uparrow$~~ & PSNR$\uparrow$ \\ \hline\hline
                    MoDL-PDFS &0.600$\pm$0.06 &29.2$\pm$1.52 &0.553$\pm$0.06 &28.2$\pm$1.15\\ 
                    Xiang \textit{et al.} &0.621$\pm$0.05 &29.4$\pm$1.71 &0.567$\pm$0.03 &28.6$\pm$1.10 \\ 
                    DuDoRNet &{\color{blue}0.641$\pm$0.05} &{\color{blue}29.6$\pm$1.39} &{\color{blue}0.577$\pm$0.05} &{\color{blue}28.8$\pm$1.29} \\ \hline
                    \textbf{MARIO}  &{\color{red} 0.646$\pm$0.03} &{\color{red}30.4 $\pm$1.22} &{\color{red} 0.588$\pm$0.04} &{\color{red} 29.6$\pm$1.19} \\ \hline
	        	\end{tabular}
			}
		\end{threeparttable}
		\vspace*{2pt}

	\end{minipage}

\end{figure*}

For visual comparison, we depict the reconstructed T2 weight images and corresponding error maps under 1D Uniform and Random patterns with 3$\times$ and 6$\times$ acceleration rates in~\figref{1D}, and the reconstructed T2 weight images and corresponding error maps under 2D Radial and Spiral patterns with 6$\times$ and 9$\times$ acceleration rates in~\figref{2D}. In general, 2D undersampling masks provide better results in comparison to 1D undersampling masks. It is obvious that the restoration of ZF creates severe aliasing artifacts and fake details. MoDL can improve the reconstruction as compared to ZF, but it still produces poor results than other T1 weight fusion methods. Our reconstructions exhibit fewer aliasing artifacts and structural losses as compared to other methods. More importantly, our method is robust under various undersampling patterns and acceleration rates, achieving better results and better-preserved anatomical details, especially at a high acceleration rate. 

\subsection{PDFS image reconstruction with the guidance of PD feature.}

\begin{figure*}[!t]
\centering
  \includegraphics[width=1\textwidth]{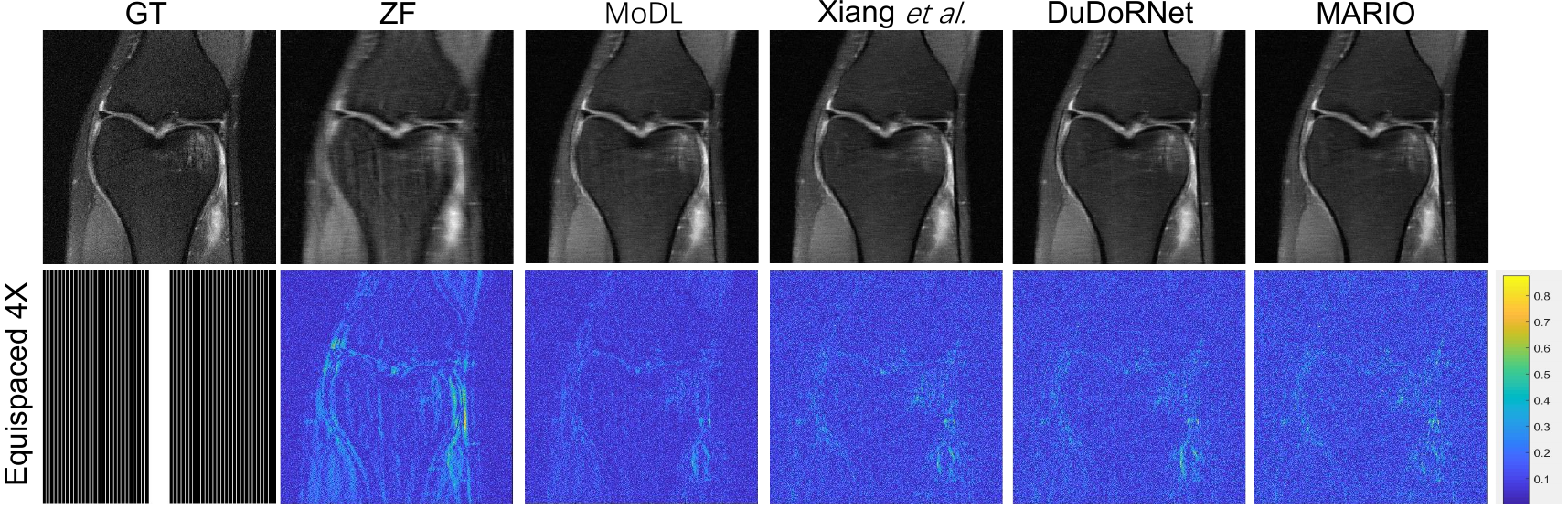}
  \caption{\textbf{ Visual comparison of different methods  on fastMRI under \textbf{Equispaced undersampling patterns} with \textbf{4$\times$} acceleration rate.} The less texture in the error map, the better the image quality.}
  \label{fastMRI} 
\end{figure*}

In the second experiment, we evaluate the reconstruction of PDFS images under the guidance of PD features. The fastMRI dataset\cite{zbontar2018fastmri} is employed, where the PDs are used as the auxiliary modality to guide the reconstruction of the target modality PDFSs\cite{xuan2020learning}.
We filter out 8332 and 1665 pairs of PD and PDFS slices for training and validation, respectively\cite{xuan2020learning}. For the fastMRI dataset, the model is trained on complex images with real and imaginary channel outputs.  In this experiment, we use the 1D Random and Equispaced sampling patterns with different acceleration rates which is the default undersampling pattern provided by the dataset to evaluate our method (shown in~\figref{fastMRI}).

The results are shown in the last two sub-tables in Table~\ref{table:1}.
MoDL without fusion of PD features can achieve the fine reconstruction result, but it is hard to see a significant improvement when a high undersampling rate is presented.  However, our method still outperforms previous methods under extremely challenging settings. This can be attributed to the powerful capability of our multi-scale fusion model both in the frequency and image domain. More importantly, the overall results in Table~\ref{table:1} reveal the strong robustness of our multi-scale fusion model under various undersampling patterns and acceleration rates.

We display the reconstructed PDFS images and corresponding error maps under Equispaced patterns with 4$\times$ acceleration rates in~\figref{fastMRI}. As shown in the figure, the images reconstructed by our method have fewer reconstruction errors and clearer structures than those produced by other methods. The superior performance is owed to the fact that our method can effectively aggregate the features of PD and PDFS images in various scales. It shows that our method provides clear anatomical details and fewer aliasing artifacts.

\begin{figure*}[!t]
\centering
  \includegraphics[width=1\textwidth]{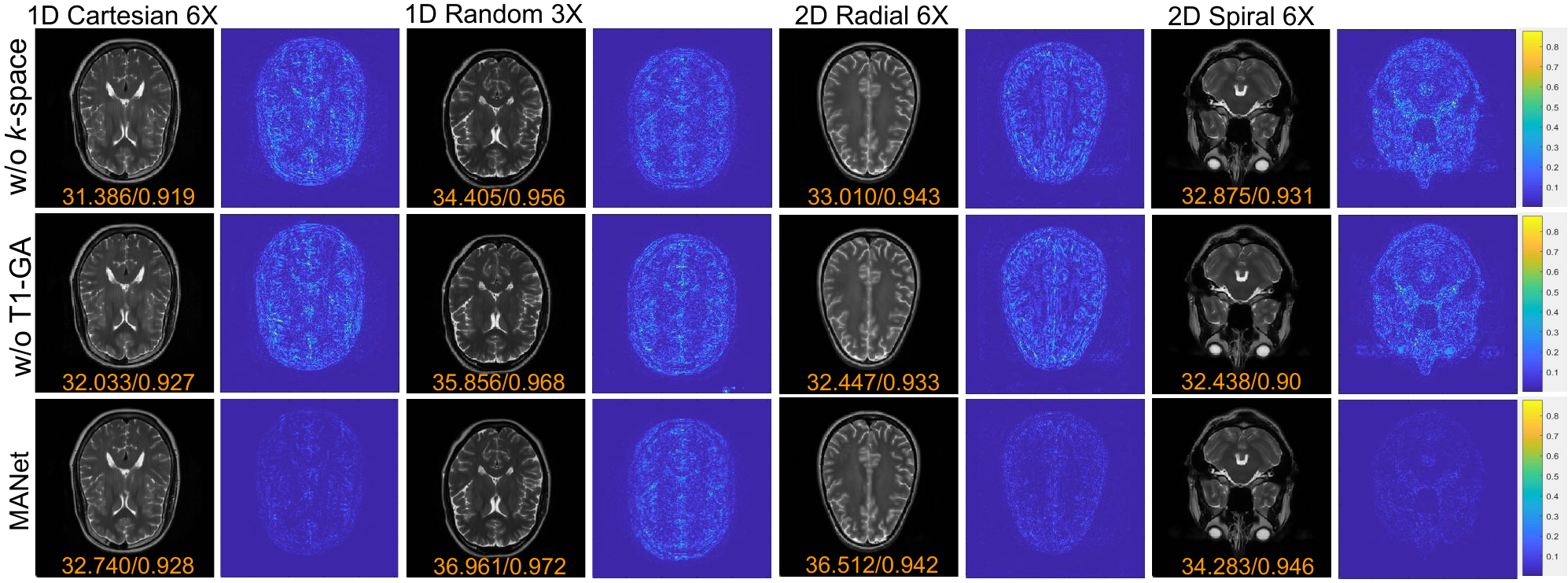}
  \caption{\textbf{Reconstruction results from 1D Uniform \textbf{6$\times$}, 1D Random \textbf{3$\times$}, 2D Radial \textbf{6$\times$} and 2D Spiral \textbf{\textbf{6$\times$}}.} Reconstruction results and error maps are presented with corresponding quantitative measurements in PSNR/SSIM. The less texture in the error map, the better the image quality.}
  \label{ablation}
\end{figure*}


\begin{figure*}[!t]
	\makeatletter\def\@captype{table}\makeatother\caption{\textbf{ Discussion on the IXI dataset under different undersampling patterns, in terms of SSIM, and PSNR.} The best and second-best results are marked in {\color{red}red} and {\color{blue}blue}, respectively. Compared with the other methods, MARIO yields the highest PSNR and SSIM results.} 
	\vspace{-5pt}
	\begin{minipage}[t]{0.5\textwidth}
		\centering
        \label{table:ab1}
		\begin{threeparttable}
			\resizebox{1\textwidth}{!}{
				\setlength\tabcolsep{4pt}
				\renewcommand\arraystretch{1.05}
				\begin{tabular}{r||cc|cc}
			        \hline\hline
			        \rowcolor{mygray}
			        &  \multicolumn{2}{c|}{1D-Uniform-6$\times$} &  \multicolumn{2}{c}{1D-Uniform-9$\times$} \\
			        \rowcolor{mygray}

			        \multirow{-2}{*}{Method}  &~~SSIM$\uparrow$~~ & PSNR$\uparrow$ &~~SSIM$\uparrow$~~ & PSNR$\uparrow$ \\ \hline\hline 
                    w/o $k$-space  &{\color{blue}0.937$\pm$0.02} &{\color{blue}32.1$\pm$1.04}  &{\color{blue}0.934}$\pm$0.15 &31.5$\pm$1.46 \\ 
                    w/o T1-GA  &0.935$\pm$0.05 &30.2$\pm$1.26  &{\color{blue}0.934$\pm$0.04} &{\color{blue}31.8$\pm$1.37} \\ \hline
                    \textbf{MARIO}   &{\color{red}0.944$\pm$0.02} &{\color{red}32.9$\pm$1.02}  &{\color{red}0.947$\pm$0.04} &{\color{red}32.4$\pm$1.16} \\ \hline
	        	\end{tabular}
			}

		\end{threeparttable}
		\vspace*{2pt}
	\end{minipage}
	\begin{minipage}[t]{0.5\textwidth}
		\centering
		\strut\vspace*{-\baselineskip}\newline\newline
		\begin{threeparttable}
			\resizebox{1\textwidth}{!}{
				\setlength\tabcolsep{4pt}
				\renewcommand\arraystretch{1.05}
				\begin{tabular}{r||cc|cc}
			        \hline\hline
			        \rowcolor{mygray}
			        &  \multicolumn{2}{c|}{1D-Random-3$\times$} &  \multicolumn{2}{c}{1D-Random-6$\times$} \\
			        \rowcolor{mygray}

			        \multirow{-2}{*}{Method}  &~~SSIM$\uparrow$~~ & PSNR$\uparrow$ &~~SSIM$\uparrow$~~ & PSNR$\uparrow$ \\ \hline\hline
                    w/o $k$-space  &0.961$\pm$0.07 &34.2$\pm$1.33  &0.926$\pm$0.06 &31.3$\pm$1.40 \\ 
                    w/o T1-GA  &{\color{blue}0.965$\pm$0.06} &{\color{blue}36.5$\pm$1.37} &{\color{blue}0.935$\pm$0.02} &{\color{blue}32.8$\pm$1.04} \\ \hline
                    \textbf{MARIO}  &{\color{red}0.969$\pm$0.06} &{\color{red}36.9$\pm$1.18} &{\color{red}0.949$\pm$0.04} &{\color{red}33.1$\pm$1.02} \\ \hline
	        	\end{tabular}
			}
		\end{threeparttable}
		\vspace*{2pt}

	\end{minipage}

	\begin{minipage}[t]{0.5\textwidth}
		\centering
		\begin{threeparttable}
			\resizebox{1\textwidth}{!}{
				\setlength\tabcolsep{4pt}
				\renewcommand\arraystretch{1.05}
				\begin{tabular}{r||cc|cc}
			        \hline\hline
			        \rowcolor{mygray}
			        &  \multicolumn{2}{c|}{2D-Radial-6$\times$} &  \multicolumn{2}{c}{2D-Radial-9$\times$} \\
			        \rowcolor{mygray}

			        \multirow{-2}{*}{Method}  &~~SSIM$\uparrow$~~ & PSNR$\uparrow$ &~~SSIM$\uparrow$~~ & PSNR$\uparrow$ \\ \hline\hline
                    w/o $k$-space &0.903$\pm$0.07 &31.1$\pm$1.54 &{\color{blue}0.915$\pm$0.07} &{\color{blue}30.8$\pm$1.50} \\ 
                    w/o T1-GA  &{\color{blue}0.930$\pm$0.07} &{\color{blue}32.6$\pm$1.49} &0.904$\pm$0.06 &28.9$\pm$1.53 \\ \hline
                    \textbf{MARIO}   &{\color{red}0.947$\pm$0.07} &{\color{red}34.3$\pm$1.31} &{\color{red}0.927$\pm$0.05} &{\color{red}32.9$\pm$1.11} \\ \hline
	        	\end{tabular}
			}
		\end{threeparttable}
		\end{minipage}
	\begin{minipage}[t]{0.5\textwidth}
		\centering
		\begin{threeparttable}
			\resizebox{1\textwidth}{!}{
				\setlength\tabcolsep{4pt}
				\renewcommand\arraystretch{1.05}
				\begin{tabular}{r||cc|cc}
			        \hline\hline
			        \rowcolor{mygray}
			        &  \multicolumn{2}{c|}{2D-Spiral-6$\times$} &  \multicolumn{2}{c}{2D-Spiral-9$\times$} \\
			        \rowcolor{mygray}

			        \multirow{-2}{*}{Method} &~~SSIM$\uparrow$~~ & PSNR$\uparrow$ &~~SSIM$\uparrow$~~ & PSNR$\uparrow$ \\ \hline\hline
                    w/o $k$-space &0.911$\pm$0.04 &30.1$\pm$1.22 &{\color{blue}0.934$\pm$0.03} &30.2$\pm$1.10 \\ 
                    w/o T1-GA  &{\color{blue}0.950$\pm$0.04} &{\color{blue}34.1$\pm$1.13} &{\color{blue}0.934$\pm$0.03} &{\color{blue}30.3$\pm$1.15} \\ \hline
                    \textbf{MARIO}  &{\color{red}0.951$\pm$0.04} &{\color{red}34.5$\pm$1.17} &{\color{red}0.940$\pm$0.03} &{\color{red}33.0$\pm$1.20} \\ \hline
	        	\end{tabular}
			}
		\end{threeparttable}
	\end{minipage}
\end{figure*}

\subsection{Ablation study to evaluate the effectiveness of each essential component in MARIO.}
In this experiment, we design two baseline models: w/o $k$-space which only employs image-domain data for reconstruction, as well as w/o T1-GA which directly concatenates the features of the two modalities rather than aggregation using the attention module. Without loss of generality, the reconstruction performance is evaluated on 1D Uniform, 1D Random, 2D Radial, 2D Spiral patterns with acceleration rates of 3$\times$, 6$\times$, and 9$\times$. The results are summarized in Table~\ref{table:ab1}. From the table, we observe that w/o $k$-space and w/o T1-GA are worse than our complete model. This indicates the importance of $k$-space information for the reconstruction, and that learning from two domains is necessary to achieve complementary representations, even when using simple fusion strategies. More importantly, MARIO further improve the results of w/o T1-GA and w/o $k$-space on all undersampling patterns and acceleration rates, showing the powerful capability of our T1-guided attention module in selectively discovering crucial information to guide T2 weighted image reconstruction.

For qualitative analysis, we visually show the reconstruction results with error maps in \figref{ablation}. It obvious that our MARIO can restore the image details with minimum errors for all undersampling patterns and acceleration rates. Combining the T1-guided attention module and hybrid domain learning, our model leads to the highest boost in performance. Our MARIO, equipping with all components, produces the best reconstruction results with minimum error. Notably, on the Spiral pattern with 6$\times$ acceleration rate, the reconstructed image of our full model is extremely high-quality, even in comparison with ground truth images.

\section*{Discussion}

Multi-modal fusion has demonstrated effectiveness in a variety of computer vision  tasks\cite{Baltrusaitis2019,Zhang2020Multimodal,Korot2021NMI}. Many natural applications can be well learned by multi-modal fusion because of the multi-modal properties of the data.
Since multi-modal learning employs the complementarity between multi-modal data, fusing various modalities has attracted increased attention recently. The advantage of multi-modal fusion is that, by learning to combine multiple modalities in a new space, it is more robust than using input features from any single modality.
Moreover, attention mechanisms allow the network to selectively pay
attention to a subset of inputs, making the network intentionally focus on useful information to improve performance. Therefore, attention mechanisms have been widely explored in visual tasks\cite{NIPS2017_3f5ee243,Wang_2017_CVPR,Hu_2018_CVPR}. Inspired by the success of attention mechanisms in visual tasks, our attention module is used to selectively enhance the features of auxiliary modality, which are crucial to target images. In our MARIO, we introduce a deep multi-modal fusion method for accelerated MR image reconstruction with the complementary of the auxiliary modality, which helps to effectively explore the correlations between multi-modal images. Based on this, we design a T1-guided attention module for multi-scale fusion, which deeply fuses the two weighted images at multiple scales, yielding high-fidelity reconstructions.

Collecting multiple MR image modalities has become the standard of clinical diagnosis at MR imaging. It is impractical to keep patients in the scanner for long periods to obtain clean and clear MR images of multiple modalities. How to accelerate MRI imaging by using supplementary information between multiple modes is a problem to be solved at present. How to use the complementary information between multiple modalities to speed up MR imaging is an open problem at present. Our MARIO can effectively mine the depth relationship between different modalities, and help the target modality for accelerated imaging through the auxiliary one with fast imaging speed. Compared with the classical reconstruction method, the prominent effect of MARIO provides a new way for the fast MR imaging system. Specifically, extensive experiments have been conducted on open datasets of MR image reconstructions. MARIO achieves better results than various single-modal and multi-modal fusion baselines. In addition, our method not only brings up high quantitative numerical results, but also has a better visual effect than other methods, which can be clearly observed by the textures in error maps (shown in~\figref{1D}, ~\figref{2D}, and ~\figref{fastMRI}). Therefore, we believe that our proposed framework is a cost-effective MR image reconstruction scheme that can aggregate multi-modal images and speed up MR imaging. 

We also observe that there are still have several limitations in our study. First, the deep convolutional neural network can only handle all spatial pixels of the image identically and cannot discriminate the target region from the background region, limiting the expression ability of the image. In future research, we'll focus on more information-rich regions and structures in MR images in order to develop faster imaging systems with improved reconstruction accuracy. In addition, although this work only considered MR multi-modal assisted imaging, MARIO may also be suitable for other medical imaging tasks, \ie, CT, X-ray, etc. The flexibility of MARIO in this area will also be evaluated in future work. 

In summary, our MARIO is an effective scheme to meet the current requirements of multi-modal MR image acquisition. MARIO provides an alternative perspective for MR image reconstruction by using data from an auxiliary modality to assist the reconstruction of the target modality, reducing overall online acquisition time and boosting patient satisfaction. On the real datasets, MARIO consistently outperforms under various undersampling patterns and acceleration rates. Notably, MARIO can still create visual effects that are close to ground truth at 6 and 9 $\times$ acceleration rates. As a result, MARIO successfully mines the internal relationships of multi-modal MR images, allowing users to speed up the reconstruction of target modalities with the easy-to-obtain modalities as supplementary information. MARIO is a multi-modal fusion deep learning model that can be trained offline and then implanted into the scanner, significantly reducing the online imaging time. Our system has the potential to improve MR imaging workflow in a cost-effective and efficient manner.

\begin{methods}
\label{sec:methods}
\subsection{Problem definition for MRI reconstruction.}
Let $\bm x_{f}$ represent the fully sampled image. We aim to reconstruct
a fully sampled image $\bm x'_{f}$ from undersampled
$k$-space $\bm k_{u}$ or undersampled
image $\bm x_{u}$. The relationship between $\bm x_{u}$ and $\bm k_{u}$ is as follows:
\begin{equation}\label{eq:relation}
\bm{x}_{u}=\mathcal{F}^{-1}\left(\bm k_{u}\right)+\epsilon=\mathcal{F}^{-1}\left(M \odot \mathcal{F}\left(\bm x_{f}\right)\right)+\epsilon,
\end{equation}
where $\mathcal{F}$ and $\mathcal{F}^{-1}$ represent the 2D Fast Fourier Transform (FFT) and inverse FFT, $M$ represents a binary mask operator to select a subset of the $k$-space points, $\epsilon$ is the noise and $\odot$ is the element-wise multiplication operation. It is worth noting that MR images obtained by $\mathcal{F}^{-1}$ are complex-valued matrices. We can reconstruct $\bm x'_{f}$ with prior knowledge of its properties, which can be formulated as the following optimization problem:
\begin{equation}
\bm x'_{f}=\underset{\bm x_{f}}{\arg\min }\|M \odot \mathcal{F} (\bm x_{f})-\bm k_{u}\|_{2}^{2}+\sum_{i} \lambda_{i} \psi_{i}(\bm x_{f}),
\label{eq:cs}
\end{equation}
where $\psi$ is a regularization function and $\lambda_i$ is a weight controlling the trade-off between the regularization terms
and data fidelity.

The ill-posed inverse problem in Eq.~\eqref{eq:cs} of reconstructing $\bm x'_{f}$ from limited sampled data can be effectively resolved by CNNs. Formally, Eq.~\eqref{eq:cs} can be well approximated using neural networks by minimizing the following loss function:
\begin{equation}\label{eq:cnn}
    \mathcal{L} = \frac{1}{N} \sum_{n=1}^N \mathcal{L}_{\text{Image}}( \mathcal{G}_\theta(\bm{x}_u^n), \bm{x}_f^n).
\end{equation}
Here, given $N$ training samples $\{\bm{x}_u^n, \bm{x}_f^n\}_{n=1}^N$, a neural network $\mathcal{G}_\theta$, parameterized by $\theta$, is learnt to reconstruct $\bm{x}_f$ from $\bm{x}_u$. $\mathcal{L}_{\text{Image}}$ denotes the loss function for measuring the image-level reconstruction, for which a variety of losses (\eg, $l_1$, $l_2$ or SSIM losses) can be used.




\subsection{MR image reconstruction via multi-modal fusion.}\label{sec:problem}
Rather than conducting reconstruction within a single modality, like in Eq.~\eqref{eq:cnn}, our motivation is to learn correlation knowledge between two closely related modalities, and employ the knowledge to improve the reconstruction. In particular, we use T1 weighted images as guidance for more accurate reconstruction of T2 weighted images. Thus, Eq.~\eqref{eq:cnn} becomes:
\begin{equation}\label{eq:cnn2}
    \mathcal{L} = \frac{1}{N} \sum_{n=1}^N \mathcal{L}_{\text{Image}}( \mathcal{G}_\theta(\bm{x}_{u,T_2}^n, \bm{x}_{f, T_1}^n), \bm{x}_{f,T_2}^n),
\end{equation}
where $\bm x_{f,T_2}$ is the fully-sampled T2 weighted image, $\bm x_{u,T_2}$ is the undersampled T2 weighted image, and $\bm x_{f,T_1}$ is the fully-sampled T1 weighted image. It is important to note that the T1 weighted images are mined as multi-scale features and fused into our framework. We aim to obtain a series of well-trained weights which are guided by the features of fully-sampled T1 weighted images. 

\subsection{Hybrid domain learning framework of MARIO.}\label{sec:hdl}
Our MARIO adopts a hybrid domain learning framework to learn comprehensive feature representations from data in both $k$-space and image domains. As illustrated in~\figref{T1T2}.\!~{\textbf{a}}, the framework consists of two sequential stages. The first stage recovers signals from the undersampled \textit{k}-space measurements and its outputs are transformed into the image domain for the second-stage image-aware reconstruction. In each stage, the corresponding T1 signal is employed to guide the reconstruction procedure.

More specifically, instead of only considering image-domain optimization as in Eq.~\eqref{eq:cnn2}, the framework optimizes a dual-domain (\ie, \textit{k}-space and image domain) learning objective, as follows:
\begin{equation}\label{eq:cnn3}
\begin{aligned}
    \mathcal{L} = \frac{1}{N} \sum_{n=1}^N (\mathcal{L}_{k\text{-space}}( \mathcal{G}_k(\bm{k}_{u,T_2}^n, \bm{k}_{f, T_1}^n), \bm{k}_{f,T_2}^n) +
    \mathcal{L}_{\text{Image}}( \mathcal{G}_\theta(\bm{x}_{u,T_2}^n, \bm{x}_{f, T_1}^n), \bm{x}_{f,T_2}^n)),
\end{aligned}
\end{equation}
where $\mathcal{G}_k$ denotes the neural network for $k$-space reconstruction, and $\mathcal{L}_{k\text{-space}}$ is the corresponding loss function for measuring the results. In our network implementation, Eq.~\eqref{eq:cnn3} is solved in the following manner (see~\figref{T1T2}.\!~{\textbf{a}}):
\begin{equation}\label{eq:cnn4}
\begin{aligned}
    \mathcal{L} =\frac{1}{N} \sum_{n=1}^N \alpha\left((\!1\!-\!\beta)\mathcal{L}_{k\text{-space}}^{T_1} + \beta\mathcal{L}_{k\text{-space}}^{T_2}\right) +
    (\!1\!-\!\alpha)\left((\!1-\!\!\beta)\mathcal{L}_{\text{Image}}^{T_1} + \beta\mathcal{L}_{\text{Image}}^{T_2}\right),
\end{aligned}
\end{equation}
where $\mathcal{L}_{k\text{-space}}^{T_2}$ and $\mathcal{L}_{\text{Image}}^{T_2}$ denote the loss functions for the target T2 modality in the $k$-space and image domains, respectively. They are similar to $\mathcal{L}_{k\text{-space}}^{T_1}$ and $\mathcal{L}_{\text{Image}}^{T_1}$. $\alpha$ weight the trade-off between the image domain and $k$-space domain, and $\beta$ weight the trade-off between the  reconstruction of the two modalities. We apply a loss on the T1 weight to enable clear structural information to be preserved, and to guide the reconstruction of the target modality. The $\alpha$s balance the weights of these losses. Eqs.~(\ref{eq:cnn3}-\ref{eq:cnn4}) finally provide our MARIO a multi-domain, multi-modal learning mechanism for reconstruction.

Our MARIO takes the same network structure for learning in both the $k$-space and image domains. Specifically, two Dense-UNets are employed for feature extraction for the T1 and T2 inputs, while a T1-guided attention module (T1-GA) is incorporated into each convolutional stage for guided learning, as shown in~\figref{T1T2}.\!~{\textbf{c}}. In the following, we describe the details of the T1-guided attention module for multi-modal aggregation.

\subsection{T1-Guided Attention Module.}\label{sec:fusion}
To ensure that the T1 weighted features can effectively guide the T2 weighted image reconstruction, we must fuse these two different modalities. Besides, any or redundancy and information that is not essential to both modalities must be reduced. Inspired by Woo \textit{et al.}\cite{woo2018cbam}, we design a T1-guided attention module to improve the compatibility of the two different modalities and excavate fused features for T2 weighted image reconstruction. The detailed architecture of the module is depicted in \figref{T1T2}.\!~{\textbf{a}}.

In particular, we devise a novel T1-guided attention module $\mathcal{M}$ to enrich the T2 weighted features. We denote $\bm{f}_{T_1}\!\in\!\mathbb{R}^{H\!\times\!W\!\times\!C}$ and $\bm{f}_{T_2}\!\in\!\mathbb{R}^{H\!\times\!W\!\times\!C}$ as the convolutional features of the T1 and T2 weighted images, respectively. The attention module can be formulated as:
\begin{equation}
\hat{\bm{f}}_{T_2}=\mathcal{M}(\bm{f}_{T_1},\bm{f}_{T_2}) \in \mathbb{R}^{H\!\times\!W\!\times\!C},
\end{equation}
where $\hat{\bm{f}}_{T_2}$ indicates the enhanced feature of the T2 modality, which is used as the input of the next convolutional block. The attention module $\mathcal{M}$ is achieved in two stages:

\textit{1) T1 Modulation:}
In this stage, we aim to modulate T1 weighted images with powerful attention modules to obtain more informative feature representation. This is achieved with channel and spatial attention modules, organized in a sequential manner (see \figref{T1T2}).
This is designed based on the famous Squeeze-and-Excitation Network (SENet)\cite{hu2018squeeze}, which does not rely on extra inputs and is trivial in computational cost. This is because the channel attention aims to reduce the dimensions, while the spatial attention aims to determine the most important information to guide the restoration of the target modality. Thus, if we place the spatial attention before the channel attention, there will be a loss of information\cite{roy2018recalibrating}.To avoid this, we organize them sequentially as the channel-spatial attention. It re-weights the feature response across channels using the \textit{squeeze} and \textit{excitation} operations. More specifically, the \textit{squeeze} operator acquires global information by aggregating the features across all the spatial locations through channel-wise global average pooling:
\begin{equation}
    \bm{z} = P_{\text{max}}(\bm{f}_{T_1}) \in \mathbb{R}^C,
\end{equation}
where $P_{\text{max}}$ indicates the global max pooling operation, and $\mathbf{z}$ is a global statistic. In the \textit{excitation} step, a gating mechanism is employed on the channel-wise descriptor:
\begin{equation}\label{eq:excite}
\bm{s}  =\sigma(\textbf{W}_{2} \delta(\textbf{W}_{1} \bm{z})) \in [0,1]^C,
\end{equation}
where $\sigma$ and $\delta$ are \textit{Sigmoid} and \textit{ReLU} functions, respectively. $\textbf{W}_{1}\!\in\!\mathbb{R}^{\frac{C}{d}\!\times\!C}$ and $\textbf{W}_{2}\!\in\!\mathbb{R}^{C\!\times\!\frac{C}{d}}$ are two fully-connected (FC) layers with $r>1$ as the reduction ratio for computational efficiency. Through Eq.~\eqref{eq:excite}, we obtain an attention vector $\bm{s}$ that encodes non-mutually-exclusive relations  among the channels of $\bm{f}_{T_1}$. The $\bm{s}$ is then applied to re-weight the channels of the original feature $\bm{f}_{T_1}$:
\begin{equation}
    \bm{f}_{T_1}^c = [\bm{s}^1 \cdot \bm{f}_{T_1}^1, \ldots, \bm{s}^C \cdot \bm{f}_{T_1}^C],
\end{equation}
where $\bm{s}^i\!\in\!\mathbb{R}$ is the $i$-th element in $\bm{s}$, and $\bm{f}_{T_1}^i\!\in\!\mathbb{R}^{H\!\times\!W}$ indicates the feature map of $i$-th channel in $\bm{f}_{T_1}$. `[]' is a channel-wise concatenation operator.

While the channel attention focuses on highlighting `what' is meaningful in $\bm{f}_{T_1}$\cite{zeiler2014visualizing}, we further discover `where' is important in $\bm{f}_{T_1}^c$ using a spatial-wise attention module $\mathcal{M}^s$. Formally, we first obtain a spatial attention map $\bm{m}$ as follows:
\begin{equation}
    \bm{m} = \sigma(\text{Conv}(Q_{\text{max}}(\bm{f}_{T_1}^c))) \in [0,1]^{H\!\times\!W},
\end{equation}
where $Q_{\text{max}}$ represents the global max pooling along the channel axis. A $7\!\times\!7$ convolutional layer is applied to the pooled feature to learn where to emphasize or suppress. Next, we employ $\bm{m}$ to enrich the channel-attentive feature $\bm{f}_{T_1}^c$:
\begin{equation}
    \bm{f}_{T_1}^s = \bm{f}_{T_1}^c \otimes \bm{m},
\end{equation}
where $\otimes$ denotes the element-wise multiplication. Through the sequential channel and spatial attention modules, we obtain a more informative feature representation $\bm{f}_{T_1}^s$. Next, we use it to enhance the features of the T2 modality.

\noindent\textit{2) T1-Guided Fusion:}
Since T1 weighted images are used to guide the reconstruction of T2 weighted images, we use the modulated T1 weighted features as complementary information and fuse them into T2 weighted features. The T1-guided fusion module works in the following residual form:
\begin{equation}
    \hat{\bm{f}}_{T_2} = \bm{f}_{T_2} \otimes \bm{f}_{T_1}^s \oplus \bm{f}_{T_2},
\end{equation}
where $\oplus$ denotes element-wise summation.
Thanks to the fusion with T1 modulation, $\hat{\bm{f}}_{T_2}$ contains rich complementary information with emphasis on T1 weighted features. By using the T1-guided attention module at each level, we obtain the depth aggregation features.

\subsection{Implementation details.} 
We implement MARIO in PyTorch and train it on two NVIDIA Tesla V100 GPUs with 32GB of memory per card. Testing is conducted on a single NVIDIA Tesla V100 GPU with a 32GB memory. To reveal full details of our method, our implementations will be released. Our model is fully end-to-end trainable. The parameters are updated using the Adam optimizer\cite{2} with an initial learning rate 0.001. We train the model for 50 epochs in total with a mini-batch size of 8. The $\alpha$ and $\beta$ in our experiments are empirically set to 0.4, and 0.6, respectively.

\subsection{Evaluation metrics.}
We use three popular metrics to evaluate our reconstruction performance: normalized mean square error (NMSE), structural similarity (SSIM) and peak signal-to-noise ratio (PSNR)\cite{zbontar2018fastmri}. Given a reconstructed image as vector $\hat{v}$ and a reference image as $v$, NMSE and PSNR are formulated as follows:
\begin{align}
\operatorname{NMSE}(\hat{v}, v)&=\frac{\|\hat{v}-v\|_{2}^{2}}{\|v\|_{2}^{2}},\\
\operatorname{PSNR}(\hat{v}, v)&=10 \log _{10} \frac{\max (v)^{2}}{\operatorname{MSE}(\hat{v}, v)},
\end{align}
where $\|\cdot\|_{2}^{2}$ is the squared Euclidean norm, and the subtraction is performed entry-wise, $\max(v)$ represents the largest entry in the ground truth, and $\operatorname{MSE}(\hat{v}, v)$ represents the mean square error between $\hat{v}$ and $v$. NMSE evaluates structural properties of the objects in an image by exploiting the inter-dependencies among nearby pixels. Given two image patches $\hat{m}$ and $m$, we have
\begin{equation}
\operatorname{SSIM}(\hat{m}, m)=\frac{\left(2 \mu_{\hat{m}} \mu_{m}+c_{1}\right)\left(2 \sigma_{\hat{m} m}+c_{2}\right)}{\left(\mu_{\hat{m}}^{2}+\mu_{m}^{2}+c_{1}\right)\left(\sigma_{\hat{m}}^{2}+\sigma_{m}^{2}+c_{2}\right)}
\end{equation}
where $\mu_{\hat{m}}$ and $\mu_{m}$ are the average pixel intensities, $\sigma_{\hat{m} m}$ is their covariance,  $\sigma_{\hat{m}}^{2}$ and $\sigma_{m}^{2}$ are their variances, and $c1$ and $c2$ are two variables to stabilize the division. Higher values of PSNR and SSIM, lower values of NMSE indicate a better reconstruction.

\end{methods}

\begin{addendum}
	\item[Competing Interests] The authors declare that they have no
	competing financial interests.
	\item[Correspondence] Correspondence and requests for materials
	should be addressed to Yong Xu and Huazhu Fu~(email: yongxu@ymail.com, hzfu@ieee.org).
	\item[Data Availability] The dataset fastMRI can be downloaded on   \url{https://fastmri.org/leaderboards}. The dataset IXI can be downloaded on   \url{http://brain-development.org/ixi-dataset/}.
	\item[Code Availability] The source code can be publicly available from  (\url{https://github.com/chunmeifeng/MARIO}).
	\item[Author Contribution] C.M.F. and H.F. proposed the idea and do the experiments, S.L. and Y.X. analyzed the results and plotted the figures, C.M.F. wrote the manuscript, D.Z. with all the authors read and revised the manuscript.
\end{addendum}

\newpage

\section*{References}
\bibliographystyle{naturemag}
\bibliography{ref}
\end{document}